\documentstyle[12pt,epsf]{article}
\begin{document}
\baselineskip 0.7cm

\newcommand{\gsim}{ \mathop{}_{\textstyle \sim}^{\textstyle >} }
\newcommand{\lsim}{ \mathop{}_{\textstyle \sim}^{\textstyle <} }
\newcommand{\vev}[1]{ \left\langle {#1} \right\rangle }
\newcommand{\nn}{\nonumber}
\renewcommand{\thefootnote}{\fnsymbol{footnote}}
\def\tr{\mathop{\rm tr}\nolimits}
\def\Tr{\mathop{\rm Tr}\nolimits}
\def\Re{\mathop{\rm Re}\nolimits}
\def\Im{\mathop{\rm Im}\nolimits}
\setcounter{footnote}{1}

\begin{titlepage}

\begin{flushright}
UT-849\\
YITP-99-29\\
\end{flushright}

\vskip 2cm
\begin{center}
{\large \bf  A Relation on Gaugino Masses \\
             in a Supersymmetric SO(10)$_{\rm GUT} \times$SO(6)$_{\rm H}$ 
             Unified Model}
\vskip 1.2cm
Kiichi Kurosawa$^{1,2}$, Yasunori Nomura$^1$, and Koshiro Suzuki$^1$

\vskip 0.4cm

{\it $^1$Department of Physics, University of Tokyo, \\
     Tokyo 113-0033, Japan}\\
{\it $^2$Yukawa Institute for Theoretical Physics, Kyoto University\\ 
     Kyoto 606-8201, Japan}
\vskip 1.5cm

\abstract{The doublet-triplet splitting problem in supersymmetric grand
 unified theories is elegantly solved in a supersymmetric 
 SO(10)$_{\rm GUT} \times$SO(6)$_{\rm H}$ model.
 In this model, the gauginos in the supersymmetric standard model do not 
 respect the usual GUT gaugino mass relation.
 We point out that in spite of non-unified gaugino masses there is one
 nontrivial relation among gaugino masses in the model.
 Thus, it can be used to test the model in future experiments.}

\end{center}
\end{titlepage}

\renewcommand{\thefootnote}{\arabic{footnote}}
\setcounter{footnote}{0}

%
%
%
%

Supersymmetric grand unified theory (SUSY GUT) \cite{SUSY-GUT} provides
an elegant explanation for the stability of the weak scale against large 
radiative corrections and peculiar hypercharge assignment in the
standard model (SM). 
It is supported by the fact that the observed three gauge coupling
constants unify at very high energy scale, $M_{\rm G} \simeq
2\times 10^{16}$ GeV \cite{GC-Unif}, and various quark and lepton
multiplets in the SM well fit into fewer multiplets of the GUT group
such as SU(5)$_{\rm GUT}$ or SO(10)$_{\rm GUT}$.

SUSY GUT models, however, generically suffer from ``doublet-triplet
splitting problem''. 
In SUSY GUT, the Higgs doublets in the SM have their color-triplet
partners. 
The masses for these triplet Higgses should be of the order of the GUT
scale in order to ensure the stability of proton and/or successful gauge
coupling unification, while those for doublet Higgses are of the order
of the weak scale. 
This requires a severe fine-tuning between parameters in the minimal
SUSY GUT models \cite{SUSY-GUT}. 
Among several mechanisms \cite{Miss_Doub, MMD, P_MMD, SO10-SO6,
Miss_VEV, M_SO10} proposed to solve the problem, one interesting
possibility is to enlarge the gauge group to the semi-simple one, 
G$_{\rm GUT} \times$G$_{\rm H}$ \cite{P_MMD, SO10-SO6}, where the
doublet-triplet splitting problem is solved by missing partner mechanism 
\cite{Miss_Doub} without introducing large representations under the GUT
group.

As shown in Ref.~\cite{Non-Unif}, this class of models may not respect
the GUT gaugino mass relation, 
\begin{eqnarray}
  \frac{m_1}{\alpha_1} = \frac{m_2}{\alpha_2} 
      = \frac{m_3}{\alpha_3},
\label{GUT-rel}
\end{eqnarray}
which is often considered as a robust prediction of SUSY GUT.
Here, $\alpha_1$, $\alpha_2$ and $\alpha_3$ 
($m_1, m_2$ and $m_3$) represent the gauge
coupling constants (gaugino masses) for SU(3)$_C$, SU(2)$_L$ and
U(1)$_Y$, respectively.\footnote{
Here, we take the GUT normalization, $\alpha_1=\frac{5}{3}\alpha_Y$.}
In this letter, we point out that in a supersymmetric unified model 
based on a semisimple gauge group SO(10)$_{\rm GUT} \times$SO(6)$_{\rm
H}$ \cite{SO10-SO6} there is a certain relation among gaugino masses in
spite of non-unified gaugino masses, and thus it can be used to test
the model by future experiments.

Let us first review the SO(10)$_{\rm GUT} \times$SO(6)$_{\rm H}$ model
proposed in Ref.~\cite{SO10-SO6} briefly. 
We introduce eleven flavors of hyperquarks $Q_\alpha^A$
$(A=1,\cdots,11; \ \alpha=1,\cdots,6)$ which transform as vector
${\bf 6}$ representations under the hypercolor gauge group 
SO(6)$_{\rm H}$.
The first ten hyperquarks $Q_\alpha^I \ (I = 1, \cdots, 10)$ form vector 
${\bf 10}$ representations and the last one $Q^{11}_\alpha$ is a
singlet of the SO(10)$_{\rm GUT}$.
We also introduce SO(6)$_{\rm H}$-singlet chiral superfields, 
$H_I({\bf 10})$, $S_{IJ}({\bf 54})$, $A_{IJ}({\bf 45})$, 
$\phi({\bf 16})$, $\bar{\phi}({\bf 16^\star})$ and $\chi({\bf 1})$ 
$(I,J = 1,\cdots,10)$, where the numbers in parentheses denote
transformation properties under the SO(10)$_{\rm GUT}$.
All the quark and lepton superfields constitute spinor {\bf 16}
representations of the SO(10)$_{\rm GUT}$ and singlets of the
SO(6)$_{\rm H}$.

In order to forbid unwanted tree-level mass terms such as $H_I H_I$ and
$Q_\alpha^{11}Q_\alpha^{11}$ in the superpotential,
we impose an anomalous global U(1)$_A$ symmetry
\begin{eqnarray}
&&  Q_\alpha^I \rightarrow Q_\alpha^I, 
\hspace{3mm}
  Q_\alpha^{11} \rightarrow e^{-2i \xi} Q_\alpha^{11},
\hspace{3mm}
  H_I \rightarrow e^{2i \xi} H_I,
\hspace{3mm}
  S_{IJ} \rightarrow S_{IJ},
\nn\\
&&  A_{IJ} \rightarrow A_{IJ},
\hspace{3mm}
  \phi \rightarrow e^{-i \xi} \phi,
\hspace{3mm}
  \bar{\phi} \rightarrow e^{i \xi} \bar{\phi},
\hspace{3mm}
  \chi \rightarrow \chi.
\label{uone}
\end{eqnarray}
The tree-level superpotential is given by
\begin{eqnarray}
  W &=& \lambda_Q \, Q^I_\alpha Q^J_\alpha S_{IJ}
  + m_Q \, Q^I_\alpha Q^I_\alpha + h \, Q_\alpha^I Q_\alpha^{11} H_I
  + \frac{1}{2} m_S \Tr(S^2) + \frac{1}{3} \lambda_S \Tr (S^3) 
\nn\\
  && + m_A \Tr(A^2) + \lambda_A \Tr(A^2 S)
  + g_\phi (\bar{\phi} \sigma_{IJ} \phi) A_{IJ}
  + g_\chi (\bar{\phi} \phi - \mu^2) \chi.
\label{pot}
\end{eqnarray}
Classically, there is an undesired vacuum 
$\vev{S_{IJ}} = \vev{Q_\alpha^I} = 0$, in which the gauge group is not
broken down to the SM one.
However, it does not exist quantum mechanically, since if
$\vev{S_{IJ}}=0$ the low-energy physics below the scale $m_Q \neq 0$
would be effectively described by an SO(6)$_{\rm H}$ gauge theory with
one massless hyperquark $Q^{11}_\alpha$ and there is no stable SUSY vacuum
in this case $(N_f \leq N_C - 5)$ \cite{dualitySO}.

Therefore, $S_{IJ}$ must have a vacuum expectation value (VEV) and
indeed we can find the following desired SUSY vacuum which is stable
quantum mechanically.
\newfont{\bgg}{cmbx10 scaled\magstep3}
\begin{eqnarray}
&& \Big\langle S_{IJ} \Big\rangle = v
\left(
\begin{array}{c@{}c@{}c@{}c@{}c}
 \frac{3}{2} {\lower0.2ex\hbox{\bf 1}} & & & & \\
 & \frac{3}{2} {\lower0.2ex\hbox{\bf 1}} & & & \\
 & &  -{\lower0.2ex\hbox{\bf 1}}& & \\
 & & & -{\lower0.2ex\hbox{\bf 1}}& \\
 & & & & -{\lower0.2ex\hbox{\bf 1}} 
\end{array}
\right),
\hspace{5mm}
\Big\langle Q_\alpha^I \Big\rangle = v_Q
\left(
\begin{array}{c@{\ }c@{\ }c}
      \noalign{\vskip 1.5ex}
      & \smash{\lower1.2ex\hbox{\bgg O}} & \\
      \noalign{\vskip 1.5ex}
      {\lower0.2ex\hbox{\bf 1}}& & \\
      & {\lower0.2ex\hbox{\bf 1}} & \\
      & & {\lower0.2ex\hbox{\bf 1}}
\end{array}
\right);
\hspace{5mm}
{\lower0.2ex\hbox{\bf 1}} = 
\left (
\begin{array}{cc}
 1 & 0\\
 0 & 1
\end{array} 
\right),
\nn\\
&& \vev{Q_\alpha^{11}} = \Big\langle H_I \Big\rangle = 0,
\\
&& \langle A_{IJ} \rangle =
  \left(
    \begin{array}{ccccc}
      a \, i \sigma_2 & & & & \\
      & a \, i \sigma_2 & & & \\
      & & b \, i \sigma_2 & & \\
      & & & b \, i \sigma_2 & \\
      & & & & b \, i \sigma_2
    \end{array}
\right);
\hspace{5mm}
  i\sigma_2 = \left (
  \begin{array}{cc}
    0 & 1 \\
    -1 & 0
  \end{array} \right ),
\nn\\
&& \langle \phi \rangle 
= v_\phi \left(\uparrow \otimes \uparrow \otimes
  \uparrow \otimes \uparrow \otimes \uparrow \right), 
\hspace{5mm}
\langle \bar{\phi} \rangle 
= v_\phi \left(\downarrow \otimes \downarrow \otimes
  \downarrow \otimes \downarrow \otimes \downarrow \right).
\nn
\label{vacuum}
\end{eqnarray}
In this vacuum the SO(10)$_{\rm GUT} \times$SO(6)$_{\rm H}$ is
broken down to the Pati-Salam gauge group 
SO(6)$_C \times$SU(2)$_L \times$SU(2)$_R$ 
(SO(6)$_C \simeq$ SU(4)$_{\rm PS}$) \cite{Pati} by the VEVs of $S_{IJ}$
and $Q_\alpha^I$, and it is further broken down to the 
SU(3)$_C\times$SU(2)$_L\times$U(1)$_Y$ by $\vev{A_{IJ}}$, $\vev{\phi}$
and $\vev{\bar{\phi}}$.
From the $F$-flatness conditions of Eq.~(\ref{pot}),
\begin{eqnarray}
&& -5 m_S v + 2 \lambda_A (a^2-b^2) 
  - \frac{5}{2} \lambda_S v^2 + 2 \lambda_Q v_Q^2 = 0,
\nn\\
&& (\lambda_Q v - m_Q) v_Q = 0,
\nn\\
&& (2 m_A + 3 \lambda_A v) a - g_\phi v_\phi^2 = 0,
\nn\\
&& (2 m_A - 2 \lambda_A v) b - g_\phi v_\phi^2 = 0,
\\
&& g_\phi (4a+6b) v_\phi + g_\chi v_\phi \chi = 0,
\nn\\
&& g_\chi (v_\phi^2 - \mu^2) = 0,
\nn
\end{eqnarray}
we obtain the VEVs $v, v_Q, a, b, v_\phi$ and $\chi$ as
\begin{eqnarray}
v &=& \frac{m_Q}{\lambda_Q},
\nn\\
v_Q^2 &=& \frac{5}{2} m_Q m_S \lambda_Q^{-2}
     + \frac{5}{4} \lambda_S m_Q^2 \lambda_Q^{-3}
     - g_\phi^2 \mu^4 \lambda_A \lambda_Q^{-1} (C_1^{-2} - C_2^{-2}), 
\nn\\ 
a &=& g_\phi \mu^2 C_1^{-1},
\nn\\
b &=& g_\phi \mu^2 C_2^{-1}, 
\\
v_\phi^2 &=& \mu^2,
\nn\\
\chi &=& - \displaystyle 10 g_\phi^2 \mu^2 
    \left( \lambda_A m_Q \lambda_Q^{-1} + 2 m_A
    \right) \displaystyle (g_\chi C_1 C_2)^{-1},
\nn
\label{vacuumAll}
\end{eqnarray}
where
\begin{equation}
  \displaystyle C_1 \equiv 2 m_A + 3 \lambda_A m_Q \lambda_Q^{-1}, \quad
  \displaystyle C_2 \equiv 2 m_A - 2 \lambda_A m_Q \lambda_Q^{-1}.
\end{equation}
We take the Yukawa couplings $\lambda_Q, \lambda_S, \lambda_A, g_\phi,
g_\chi \sim {\cal O}(1)$ and $m_Q, m_S, m_A, \mu \sim M_{\rm G}$, which
is suggested from the renormalization group analysis on the gauge
coupling constants of the low-energy gauge groups \cite{GC-Unif}. 

In view of Eq.~(\ref{pot}), the colored Higgs $H_a \ (a = 5, \cdots,
10)$ obtain the GUT scale masses together with $Q_\alpha^{11}$ but the
Higgs $H_i \ (i = 1, \cdots, 4)$ remain massless as long as $\langle
Q_\alpha^{11} \rangle = 0$.
These massless Higgs multiplets transform as ({\bf 2}, {\bf 2}) under
the SU(2)$_L \times$SU(2)$_R$ and are identified with two Higgs doublets
in the SUSY standard model.
The masslessness of $H_i$ is guaranteed by the U(1)$_{A'}$ symmetry which
is an unbroken linear combination of the U(1)$_A$ and a U(1) subgroup of 
the SO(10)$_{\rm GUT}$.\footnote{
The U(1)$_{A'}$ symmetry also forbids dangerous dimension-five operators
\cite{dim_5} which induce the proton decay through colored-Higgs
exchanges.}
On the other hand, the mass term for the colored Higgs $H_a$ and
$Q_\alpha^{11}$ is allowed since they have the opposite U(1)$_{A'}$
charges each other.
Note that the presence of the vacuum with unbroken U(1)$_{A'}$ in
Eqs.~(\ref{vacuum}) is a dynamical consequence of the present model.

Let us now discuss the gauge coupling constants and gaugino masses at
low energies.
The SM gauge fields are linear combinations of those of the SO(10)$_{\rm
GUT}$ and the SO(6)$_{\rm H}$, so that one may wonder if the successful
gauge coupling unification is spoiled by the presence of hypercolor
gauge interactions.
However, it is not necessarily true.
Assuming that the SO(10)$_{\rm GUT} \times$SO(6)$_{\rm H}$ is broken
down to the SU(3)$_C\times$SU(2)$_L\times$U(1)$_Y$ at the GUT scale for
simplicity, the SM gauge couplings are given as
\begin{eqnarray}
\frac{1}{\alpha_3} &=&
    \frac{1}{\alpha_{\rm GUT}} + \frac{1}{\alpha_{\rm H}},
\\
\frac{1}{\alpha_2} &=&
    \frac{1}{\alpha_{\rm GUT}},
\\
\frac{1}{\alpha_1} &=&
    \frac{1}{\alpha_{\rm GUT}} + \frac{2}{5}\frac{1}{\alpha_{\rm H}},
\end{eqnarray}  
at the GUT scale.
Here, $\alpha_{\rm GUT}$ and  $\alpha_{\rm H}$ are the gauge coupling
constants of the SO(10)$_{\rm GUT}$ and the SO(6)$_{\rm H}$.
Thus, if hypercolor SO(6)$_{\rm H}$ is sufficiently strong, 
$\alpha_{\rm H}(M_{\rm G}) \gg \alpha_{\rm GUT}(M_{\rm G})$, the GUT
unification $\alpha_1(M_{\rm G}) \simeq \alpha_2(M_{\rm G}) \simeq
\alpha_3(M_{\rm G})$ is achieved naturally.\footnote{
The correction from the SO(6)$_{\rm H}$ may explain the slight
discrepancy of $\alpha_3(M_Z)$ between the experimental value and the
prediction of the minimal SUSY GUT \cite{Non-Unif}.}

The gaugino masses $m_1, m_2$ and $m_3$ are also given by linear
combinations of the gaugino masses, $m_{\rm GUT}$ and $m_{\rm H}$, for
the SO(10)$_{\rm GUT}$ and the SO(6)$_{\rm H}$ as
\begin{eqnarray}
\frac{m_3}{\alpha_3} &=&
    \frac{m_{\rm GUT}}{\alpha_{\rm GUT}} 
  + \frac{m_{\rm H}}{\alpha_{\rm H}},
\label{gaugino_1}
\\
\frac{m_2}{\alpha_2} &=&
    \frac{m_{\rm GUT}}{\alpha_{\rm GUT}},
\label{gaugino_2}
\\
\frac{m_1}{\alpha_1} &=&
    \frac{m_{\rm GUT}}{\alpha_{\rm GUT}} 
  + \frac{2}{5}\frac{m_{\rm H}}{\alpha_{\rm H}}.
\label{gaugino_3}
\end{eqnarray}
Here, we have adopted the hidden sector SUSY breaking scenario.
This shows that the GUT gaugino mass relation, Eq.~(\ref{GUT-rel}),
can be broken in general.
Note that the above equations depend only on the combinations $m/\alpha$ 
which are invariant under renormalization group at one-loop level, so
that these results hold at any scale and are independent of the breaking
scales $v, v_Q, a, b$ and $v_\phi$ at one-loop level
\cite{indep}.

Next, we discuss phenomenological implications of the above equations
Eqs.~(\ref{gaugino_1} -- \ref{gaugino_3}). 
We first consider the simplest case where the gaugino masses are
originated only from the $F$-term of a dilaton superfield.
In this case, the gaugino masses $m$ are universal for all gauge groups
($m_{\rm GUT} = m_{\rm H}$) at the cut-off scale $M_*$ (string scale or
Planck scale).
Then, since gauge coupling unification requires that 
$\alpha_{\rm H}/\alpha_{\rm GUT}(M_{\rm G}) \gg 1$, one might think that 
the deviation from the GUT gaugino mass relation, Eq.~(\ref{GUT-rel}),
is small in view of Eqs.~(\ref{gaugino_1} -- \ref{gaugino_3}).
However, $\alpha_{\rm H}/\alpha_{\rm GUT}(M_{\rm G}) \gg 1$ does not
mean $\alpha_{\rm H}/\alpha_{\rm GUT}(M_*) \gg 1$.
Indeed, in the present model the hypercolor SO(6)$_{\rm H}$ is
asymptotically free while the SO(10)$_{\rm GUT}$ is not above the GUT
scale, so that $\alpha_{\rm GUT}$ and $\alpha_{\rm H}$ can be comparable
at $M_*$.
This implies that both $m_{\rm GUT}/\alpha_{\rm GUT}$ and 
$m_{\rm H}/\alpha_{\rm H}$ can make comparable contributions to the SM
gaugino masses even in this simplest case.
Moreover, gaugino masses can be non-universal in more general cases
where gaugino masses arise from $F$-terms of several moduli fields.
Thus, we take $m_{\rm GUT}/\alpha_{\rm GUT}(M_{\rm G})$ and 
$m_{\rm H}/\alpha_{\rm H}(M_{\rm G})$ as independent parameters,
hereafter.

\begin{figure}
\epsfxsize=8cm
\centerline{\epsfbox{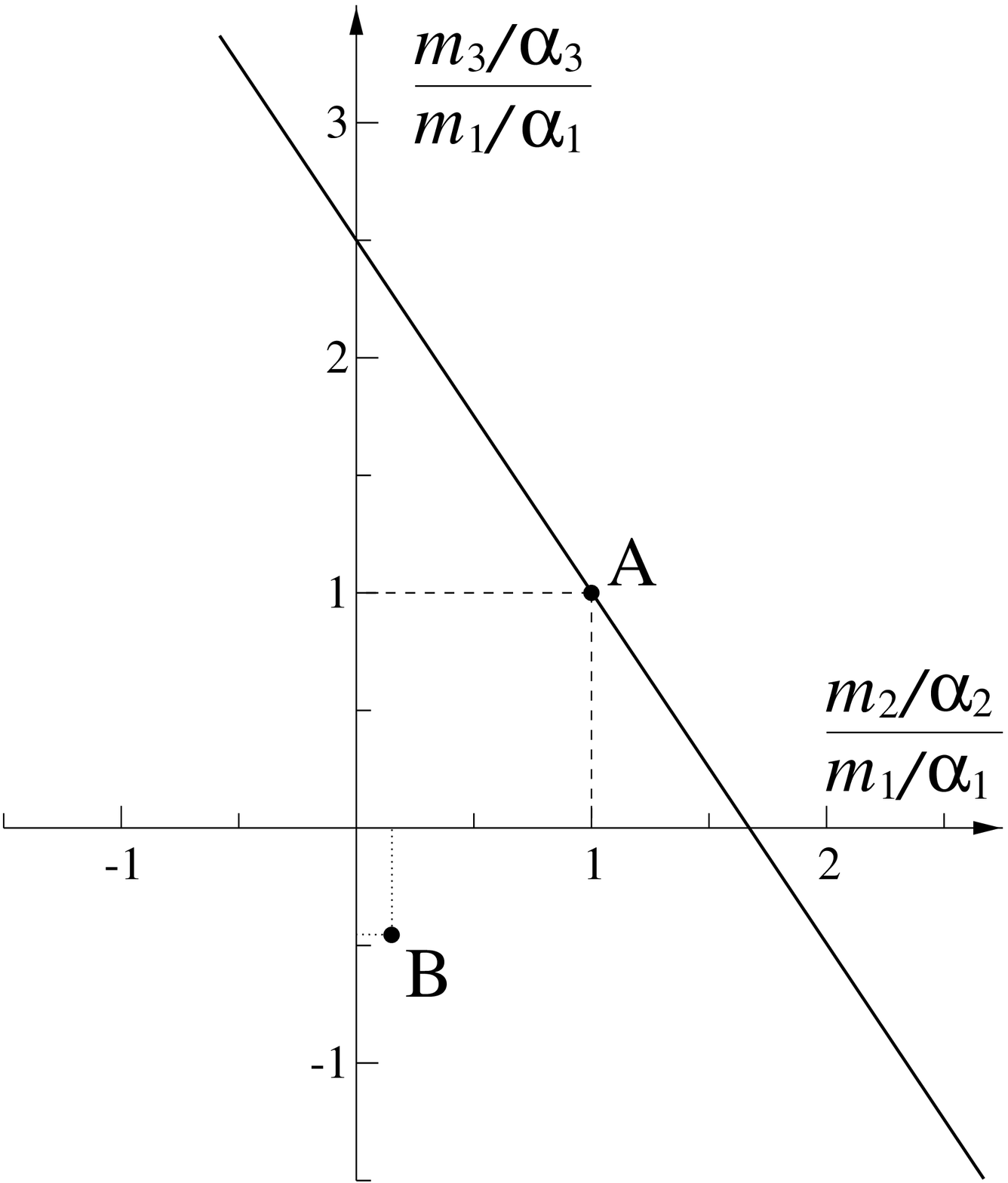}}
\caption{The gaugino mass relation predicted in the 
 SO(10)$_{\rm GUT} \times$SO(6)$_{\rm H}$ model.
 The point A denotes the case that the GUT gaugino mass relation, 
 $m_1/\alpha_1 = m_2/\alpha_2 = m_3/\alpha_3$, holds.
 The point B represents the case that the gaugino masses arise only from 
 superconformal anomaly, $m_i/\alpha_i = (b_i/4\pi) m_{3/2}$
 $(i=1,2,3)$, where $m_{3/2}$ is the gravitino mass and $b_i$ are the
 coefficients of one-loop beta functions, $(3, -1, -33/5)$, for 
 $(\alpha_3, \alpha_2, \alpha_1)$.}
\label{fig}
\end{figure}

In spite of non-unified gaugino masses, 
Eqs.~(\ref{gaugino_1} -- \ref{gaugino_3}) suggest that there is one
nontrivial relation among gaugino masses for SU(3)$_C$, SU(2)$_L$ and
U(1)$_Y$.
It is given by eliminating $m_{\rm GUT}/\alpha_{\rm GUT}$ and 
$m_{\rm H}/\alpha_{\rm H}$ as
\begin{eqnarray}
  \frac{m_1}{\alpha_1} = 
    \frac{3}{5}\frac{m_2}{\alpha_2} + \frac{2}{5}\frac{m_3}{\alpha_3}.
\label{final}
\end{eqnarray}
We stress again that this relation holds at any scale and is independent 
of symmetry breaking scales at the leading order.
We have depicted this gaugino mass relation by the solid line in
Fig.~\ref{fig}.
The horizontal and vertical axes represent $(m_2/\alpha_2)/(m_1/\alpha_1)$
and $(m_3/\alpha_3)/(m_1/\alpha_1)$, respectively.
The point A denotes the case where the GUT gaugino mass relation,
Eq.~(\ref{GUT-rel}), holds.
Here, we have assumed that $m_{\rm GUT}/m_{\rm H}$ is real in order not
to introduce SUSY CP problem.
We conclude that the gaugino masses in the present model can deviate
from those in the minimal SUSY GUT case but still maintain one relation, 
Eq.~(\ref{final}).
Note, however, that the above gaugino masses are the running gaugino
masses at one-loop level, so that the higher-loop effects and threshold
corrections which typically induce a few percent contributions
\cite{corr} should be taken into account when we compare them with the
pole masses precisely.\footnote{
The gauginos for SU(2)$_L$ and U(1)$_Y$ mix with the Higgsinos resulting 
in physical particles, charginos and neutralinos, so that $m_1$ and
$m_2$ have to be determined experimentally disentangling these
complications \cite{comp}.}

Several comments are in order.
First, the gauginos for the SM are linear combinations of those of the
SO(10)$_{\rm GUT}$ and the SO(6)$_{\rm H}$, while the squarks and
sleptons purely come from ${\bf 16}$ representations of the 
SO(10)$_{\rm GUT}$.
As a result, there may be certain sum rules for squark and slepton masses
and they can be used to determine the symmetry breaking pattern and scale
\cite{indep}.\footnote{
Realistic quark and lepton mass matrices can be obtained by
introducing appropriate nonrenormalizable interactions 
\cite{mass_ql, SO10-SO6}.}
Second, although we have assumed that $m_{\rm GUT}/m_{\rm H}$ is real in 
our analysis, $m_{\rm GUT}$ and $m_{\rm H}$ could have small relative
phases in general.\footnote{
It may be possible that this phase is even of order one \cite{O1-CP}.}
It may induce observable CP-violating effects and can be used to
discriminate the model in future experiments, since this phase cannot be 
included in the usual SUSY GUT models.
In this case, the gaugino masses will slightly deviate from 
Eq.~(\ref{final}), keeping an inequality 
$|m_1|/\alpha_1 \leq (3/5)|m_2|/\alpha_2 + (2/5)|m_3|/\alpha_3$.
Third, if gaugino masses are generated only by the SM gauge interactions
at low energies \cite{gauge_med} or by superconformal anomalies
\cite{anom_med}, the gaugino masses generically fall into point A and B
in Fig.~\ref{fig}, respectively.
In these cases, the present model is not distinguishable from the other
GUT models.\footnote{
There is a gauge mediation model which predicts the same gaugino mass
relation as in Eq.~(\ref{final}) due to accidental cancellations of
the leading order diagrams responsible for the gaugino masses
\cite{INTY}.
This case can be discriminated from the model considered in the text
by measuring squark and slepton masses.}

To summarize, we have shown that the gaugino masses can deviate
from the GUT gaugino mass relation in the SUSY 
SO(10)$_{\rm GUT} \times$SO(6)$_{\rm H}$ model.
In spite of non-unified gaugino masses, however, there is one nontrivial 
relation among the SM gaugino masses, which is independent of symmetry
breaking scales at the leading order.
Thus, observing the gaugino mass relation Eq.~(\ref{final}) in future
experiments could test the present model together with the
measurement of squark and slepton masses.

\vspace{5mm}

We would like to thank T.~Yanagida for valuable discussions 
and a careful reading of the manuscript.
Y.N. is supported by the Japan Society for the Promotion of Science.

\newpage
%
%
%
\newcommand{\Journal}[4]{{\sl #1} {\bf #2} {(#3)} {#4}}
\newcommand{\PL}{\sl Phys. Lett.}
\newcommand{\PR}{\sl Phys. Rev.}
\newcommand{\PRL}{\sl Phys. Rev. Lett.}
\newcommand{\NP}{\sl Nucl. Phys.}
\newcommand{\ZP}{\sl Z. Phys.}
\newcommand{\PTP}{\sl Prog. Theor. Phys.}
\newcommand{\NC}{\sl Nuovo Cimento}
\newcommand{\MPL}{\sl Mod. Phys. Lett.}
\newcommand{\PRep}{\sl Phys. Rep.}

\newpage

\begin{thebibliography}{99}
%
\bibitem{SUSY-GUT}
    S.~Dimopoulos and H.~Georgi,
        {\NP} {\bf B193} (1981) 150;\\
    N.~Sakai,
        {\ZP} {\bf C11} (1981) 153.
%
\bibitem{GC-Unif}
    J.~Ellis, S.~Kelley, and D.V.~Nanopoulos,
        {\PL} {\bf B260} (1991) 131;\\
    U.~Amaldi, W.~de~Boer, and H.~F\"{u}rstenau,
        {\PL} {\bf B260} (1991) 447;\\
    P.~Langacker and M.-X.~Luo,
        {\PR} {\bf D44} (1991) 817;\\
    C.~Giunti, C.W.~Kim, and U.W.~Lee,
        {\MPL} {\bf A6} (1991) 1745.
%
\bibitem{Miss_Doub}
    A.~Masiero, D.V.~Nanopoulos, K.~Tamvakis, and T.~Yanagida,
        {\PL} {\bf B115} (1982) 380;\\
    B.~Grinstein,
        {\NP} {\bf B206} (1982) 387.
%
\bibitem{MMD}
    J.~Hisano, T.~Moroi, K.~Tobe, and T.~Yanagida,
        {\PL} {\bf B342} (1995) 138,
        {\tt [hep-ph/9406417]}.
%
\bibitem{P_MMD}
    T.~Yanagida,
        {\PL} {\bf B344} (1995) 211,
        {\tt [hep-ph/9409329]};\\
    T.~Hotta, K.-I.~Izawa, and T.~Yanagida,
        {\PR} {\bf D53} (1996) 3913,
        {\tt [hep-ph/9509201]};
        {\PL} {\bf B409} (1997) 245,
        {\tt [hep-ph/9511431]};
        {\PTP} {\bf 95} (1996) 949,
        {\tt [hep-ph/9601320]}.
%
\bibitem{SO10-SO6}
    T.~Hotta, K.-I.~Izawa, and T.~Yanagida,
        {\PR} {\bf D54} (1996) 6970,
        {\tt [hep-ph/9602439]}.
%
\bibitem{Miss_VEV}
    S.~Dimopoulos and F.~Wilczek,
        in {\it The Unity of the Fundamental Interactions}, 
        ed. A.~Zichichi 
        (UC at Santa Barbara, 1981), p. 237;\\
    M.~Srednicki,
        {\NP} {\bf B202} (1982) 327.
%
\bibitem{M_SO10}
    K.S.~Babu and S.M.~Barr,
        {\PR} {\bf D48} (1993) 5354,
        {\tt [hep-ph/9306242]};
        {\PR} {\bf D50} (1994) 3529,
        {\tt [hep-ph/9402291]};\\
    J.~Hisano, H.~Murayama, and T.~Yanagida,
        {\PR} {\bf D49} (1994) 4966.
%
\bibitem{Non-Unif}
    N.~Arkani-Hamed, H.-C.~Cheng, and T.~Moroi,
        {\PL} {\bf B387} (1996) 529,
        {\tt [hep-ph/9607463]}.
%
\bibitem{dualitySO}
    K.~Intriligator and N.~Seiberg,
        {\NP} {\bf B444} (1995) 125, 
        {\tt [hep-th/9503179]}.
%
\bibitem{Pati}
    J.C.~Pati and A.~Salam,
        {\PR} {\bf D10} (1974) 275.  
%
\bibitem{dim_5}
    N.~Sakai and T.~Yanagida,
        {\NP} {\bf B197} (1982) 533;\\  
    S.~Weinberg,
        {\PR} {\bf D26} (1982) 287.  
%
\bibitem{indep}
    Y.~Kawamura, H.~Murayama, and M.~Yamaguchi,
        {\PL} {\bf B324} (1994) 52,
        {\tt [hep-ph/9402254]};
        {\PR} {\bf D51} (1995) 1337,
        {\tt [hep-ph/9406245]}.
%
\bibitem{corr}
    G.D.~Kribs,
        {\NP} {\bf B535} (1998) 41,
        {\tt [hep-ph/9803259]}.
%
\bibitem{comp}
    T.~Tsukamoto, K.~Fujii, H.~Murayama, M.~Yamaguchi, and Y.~Okada,
        {\PR} {\bf D51} (1995) 3153;\\
    J.L.~Feng and M.J.~Strassler,
        {\PR} {\bf D51} (1995) 4661,
        {\tt [hep-ph/9408359]};
        {\PR} {\bf D55} (1997) 1326,
        {\tt [hep-ph/9606477]};\\
    H.~Baer, R.~Munroe, and X.~Tata,
        {\PR} {\bf D54} (1996) 6735,
        {\sl Erratum-ibid.} {\bf D56} (1997) 4424,
        {\tt [hep-ph/9606325]}.
%
\bibitem{mass_ql}
    G.~Anderson, S.~Raby, S.~Dimopoulos, L.J.~Hall, and G.D.~Starkman,
        {\PR} {\bf D49} (1994) 3660,
        {\tt [hep-ph/9308333]}.
%
\bibitem{O1-CP}
    T.~Ibrahim and P.~Nath,
        {\PR} {\bf D58} (1998) 111301,
        {\tt [hep-ph/9807501]};\\
    M.~Brhlik, G.J.~Good, and G.L.~Kane,
        {\PR} {\bf D59} (1999) 115004,
        {\tt [hep-ph/9810457]}.
%
\bibitem{gauge_med}
    M.~Dine, A.E.~Nelson, and Y.~Shirman,
        {\PR} {\bf D51} (1995) 1362,
        {\tt [hep-ph/9408384]};\\
    M.~Dine, A.E.~Nelson, Y.~Nir, and Y.~Shirman,
        {\PR} {\bf D53} (1996) 2658,
        {\tt [hep-ph/9507378]};\\
    For a review, see G.F.Giudice and R.~Rattazzi,
        hep-ph/9801271;\\
    S.L.~Dubovsky, D.S.Gorbunov, and S.V.~Troitsky,
        hep-ph/9905466.
%
\bibitem{anom_med}
    L.~Randall and R.~Sundrum,
        hep-th/9810155;\\
    G.F.~Giudice, M.A.~Luty, H.~Murayama, and R.~Rattazzi,
        {\sl JHEP} {\bf 9812} (1998) 027,
        {\tt [hep-ph/9810442]}.
%
\bibitem{INTY}
    K.I.~Izawa, Y.~Nomura, K.~Tobe, and T.~Yanagida,
        {\PR} {\bf D56} (1997) 2886,
        {\tt [hep-ph/9705228]};\\
    Y.~Nomura and K.~Tobe,
        {\PR} {\bf D58} (1998) 055002,
        {\tt [hep-ph/9708377]}.
\end{thebibliography}
\end{document}